\documentclass[fleqn,twoside]{article}
\usepackage{espcrc2,isolatin1}

\usepackage{epsf}
\usepackage[figuresright]{rotating}

\newcommand{\AmS}{{\protect\the\textfont2
  A\kern-.1667em\lower.5ex\hbox{M}\kern-.125emS}}

\def\fig#1{Fig.~\ref{#1}}

\newcommand{\be}{\begin{equation}}
\newcommand{\ee}{\end{equation}}
\newcommand{\bea}{\begin{eqnarray}}
\newcommand{\eea}{\end{eqnarray}}
\newcommand{\noi}{\noindent}

\renewenvironment{itemize}{\begin{list}{$\bullet$}
{\itemsep-2pt}}
{\end{list}}

\title{
{
\vspace{-3.0cm} \normalsize \hfill
\parbox{30mm}{ CPT-2002/P.4417} 
}\\[22mm]
The CKM Lattice Working Group initiative\thanks{Plenary talk by L.
Lellouch at {\em Lattice 2002}, Cambridge MA, USA.
Work 
supported in part by EU HPP contract HPRN-CT-2000-00145.
\newline
$^\ddag$ Unité Propre de Recherche 7061.}}

\author{Jonathan Flynn\address[soton]{Dept. of Physics and Astronomy, 
University of Southampton, Southampton SO17 1BJ, UK}, 
Laurent Lellouch\address[cpt]{Centre de 
Physique Théorique$^\ddag$, Case 907, CNRS
Luminy, F-13288 Marseille Cedex 9, France}
and Guido Martinelli\address[roma]{Dip. di Fisica,
Univ. di Roma ``La Sapienza'', %and INFN, Sezione di Roma
P.le A. Moro 2, I-00185 Rome, Italy}
for the CKM-LWG}

\begin{document}

\begin{abstract}
We present the CKM-LWG, a collection of working groups whose aim is to
compile world averages of lattice results relevant for
high-energy phenomenology.
\vspace{-0.5cm}
\end{abstract}

\maketitle

\section{Short history}

In February 2002, the first meeting of the Workshop on the CKM
Unitarity Triangle was held at CERN. The aim of the workshop is to
review the state of the art in CKMology and propose avenues for new
physics searches. Because lattice QCD has an important rôle to play in
this program, the meeting was viewed by the authors as an excellent
opportunity to bring together lattice theorists to begin setting up a
L(attice) D(ata) G(roup) whose goal would be to compile averages for
quantities calculated on the lattice which are relevant for
high-energy phenomenology.  Representatives from major lattice
collaborations from around the world were thus invited to attend a
first assembly on February 13 at CERN.

The response was certainly positive. The meeting was well
attended \cite{ldg} and most colleagues present shared the belief that
it would be very helpful to join forces to combine results. Many also
expressed the wish to contribute and to see something come out of this
initiative.

A number of objections were raised, however. The first was that the
project was too ambitious. Most lattice phenomenology results are
still quenched or partially quenched and are, in that sense, still
preliminary. Too close a parallel with the PDG, who compile
experimental results, should thus be avoided. Another concern was
manpower, especially if a large number of quantities are to be
reviewed all at once: an enterprise such as the LDG should not come at
the expense of progress in our field. It was also noted that we, as a
community, should build up some experience with this sort of
compilation effort before embarking on a full-fledged PDG type
enterprise. 

After some discussion, it was concluded that the project should focus
on a more manageable task and prove feasibility. It was agreed that
three test working groups should be set up to review a number
of well studied quantities:
\begin{itemize}
\item
quark masses,
\item
the kaon $B$-parameter, $B_K$,
\item
matrix elements relevant for neutral $B$-meson mixing.
\end{itemize}
It was further agreed that only results published in refereed journals
should be taken into account; that the lattice community as a whole
should be informed about the project and given a chance to react and
that wider participation should be solicited (this was the subject of
an email sent to the mailing list of this conference last spring and
is the purpose of the present talk and of the follow-on meeting);
that the results of the analyses and the process which led to
them should be presented at the first Lattice Symposium following
their release. For the time being, the working groups are known
collectively as the CKM Lattice Working Group (CKM-LWG).

\section{Why do we need a CKM-LWG?}

\begin{figure}[t]
\epsfxsize=7.5cm\epsffile{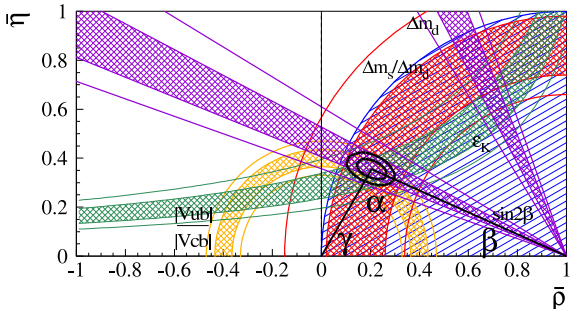}
\vspace{-1.cm}
\caption{\em Constraints on the unitarity triangle (summer 2002) 
\cite{parodiut}. 
The non-perturbative matrix elements required to constrain
the sides and summit were taken from lattice calculations.}
\vspace{-0.5cm}
\label{fig:ckm}
\end{figure}

Lattice results already play an important rôle in a number of aspects
of the current experimental endeavor in high-energy physics.  For
instance, the discovery of new physics from analyses of the CKM
unitarity triangle has come to rely more and more on weak matrix
elements obtained through lattice calculations (please see
\fig{fig:ckm}). We thus have a responsibility as a community to
provide reliable estimates of the relevant quantities and especially
of their uncertainties: an underestimate could raise false hopes of
new physics and discredit our field; an overestimate could mean that
an unique opportunity to reveal new physics is lost.

Another important aspect is that for many quantities of
phenomenological relevance, there exist a large number of calculations
using different actions, different approaches to renormalization,
etc. Furthermore, lattice theorists are curious to try out new
approaches so that more recent does not necessarily imply more
reliable or accurate. The situation can thus be very confusing for
non-experts, as it is even for experts! It would therefore be very
useful to have a body of theoreticians well versed in the different
possible lattice approaches, which could review the situation and
provide a {\em succinct} and {\em reliable} summary of the state of
the art. We also owe it to ourselves to make the best possible use of
the results and experience that we have accumulated.

\section{Isn't that the rôle of lattice rapporteurs?}

Plenary speakers at lattice conferences are mainly expected to
summarize and comment on the developments of the last year--many of
which are still preliminary--and to give their views on future
directions. Moreover, lattice plenaries can be rather technical and
phenomenological information is often scattered over different
reviews. This is to be contrasted with the focus of the CKM-LWG which
is on providing the wider particle physics community with summary
numbers for quantities of phenomenological interest which are based
solely on results published in refereed journals (i.e. old news by
lattice conference standards) and which are easily accessible.

Unlike plenary speakers at lattice symposia, the lattice rapporteur at
large particle physics conferences is usually asked to put together a
one hour or less summary of the situation in lattice field theory for
subjects ranging from chiral fermions, through finite temperature
phase transitions, to quark masses. In such a context, choices have to
be made and it is clearly very difficult to provide a summary of the
current situation for the whole range of phenomenologically relevant
quantities.  In contrast, each of the CKM Lattice Working Groups
focuses on specific quantities and is composed of the physicists who
have calculated and studied them.  The working groups therefore have
access to expertise in the different possible approaches as well as to
details of the analyses, intermediate results, directions investigated
but not retained etc. which can all be very helpful for optimally
combining results from different calculations.

To conclude on this aspect, lattice reviews and the CKM-LWG are
complementary: the CKM-LWG should make use of the expertise put into
lattice reviews and lattice reviews could use the work of the CKM-LWG
as a standard against which new developments may be compared.  A
possible scenario would be to have the CKM-LWG present its results in
winter/spring so as to best carry out this complementarity; so as to
allow time for the results presented at the lattice conference to make
it to press; so as to provide updates in time for the phenomenological
analyses presented in summer (spring?) conferences.

\section{Current status of the CKM-LWG}

The idea is to have up-to-date web pages which summarize the status of
the CKM-LWG's work and which serve as an archive for past results
\cite{ldg}. For the moment they contain the agenda of the CERN
meeting, the CKM-LWG proposal and lists of the current membership of
the three working groups. They also contain a set of suggested
guidelines, which should give a flavor of the
work required. The working groups should:
\begin{itemize}
\item
compile {\em all} published results for a given quantity, specifying
details of the calculation and systematic errors considered;

\item
determine a set of systematic errors ascribed to each quantity and
each lattice approach and use the literature to estimate their size;

\item
on the basis of this information, exclude results for which these
systematics are not under control and for the others, assign
systematic errors which take into account the analysis choices made by
authors;

\item
devise a means for combining the various results for a given
quantity, distinguishing correlated and uncorrelated systematics.
Results using similar approaches could first be combined and
the resulting numbers further compiled into a global average;

\item
consult one another on common issues.
\end{itemize}
In addition, collaborations are strongly encouraged to make available
any data which may help in the work of the different groups.

These guidelines are certainly not meant to be restrictive and each
group should organize itself and its work as it sees fit. The
experiences of the groups can then be compared and the best of each
retained. This will hopefully lead to a set of generally accepted
criteria for computing global averages, keeping in mind that different
subjects sometimes require different approaches.

Some very preliminary work has begun. Two of the working groups, for
instance, have set up a web page and some bibliographies have been
collected. However, time was too short and schedules too full to
produce even preliminary analyses for this conference.

\section{What else could the CKM-LWG do for our community?}

When presented with the idea, Karl Jansen made the pertinent remark
that another rôle of the PDG is to provide concise and accessible
reviews on subjects which are important for the every day work of
particle physicists. Following this line of thought one can imagine
collecting, say on a web page, information that would be useful to our
community (including students in our field) as well as to others
interested in understanding what we do, e.g.
\begin{itemize}

\item
descriptions of the various algorithms used, including those for
inversions, evaluating the overlap Dirac operator, etc.,

\item
estimates of the performances of these algorithms,

\item
a description of the basic fitting and analysis techniques,

\item
a summary of the different fermion and gauge actions used,

\item
plaquette values for different actions,

\item
reviews on different systematics,

\item
a review of the possible approaches to renormalization,

\item
a collection of perturbative expressions for matching factors for
different actions and quantities,

\item
a collection of non-perturbative (NP) matching factors and NP lagrangian
parameters for different actions and quantities.

\end{itemize}
The possibilities are numerous and having a unique repository for
this kind of information could be very helpful.

One could also imagine making available gauge configurations,
propagators, codes, etc. This, however, is a different business and a
very interesting proposal along these lines \cite{richard}, which
advocates the creation of an ``International Lattice Data Grid''
(ILDG), was presented at this conference
\cite{christine}.

\section{Conclusion}

A number of us believe that an initiative such as the CKM-LWG would be
very useful for high-energy physics as well as for the lattice
community. We are currently running a feasibility experiment with
three working groups on:
\begin{itemize}
\item
quark masses,
\item
$B_K$,
\item
$B-\bar B$ mixing,
\end{itemize}
and plan to produce a first CKM-LWG summary by the Lattice 2003
conference in Tsukuba at the latest. At that point, the experiences of
each group will be compared, and decisions on where to go from there
will be made.

The CKM-LWG, and perhaps later the LDG, can also serve our community
by putting together information which is useful to our everyday work
(reviews on algorithms, etc.). Work on this aspect could 
start now, though it should not divert (too many) resources from the
phenomenological missions of the CKM-LWG.

In sum, possibilities are numerous and it is up to us to decide what
we might want an LDG to be. Anyone interested in joining can contact
the authors or any other member of the CKM-LWG.

\bigskip

{\bf\noi Note added}

\medskip

In the evening following this talk, an open meeting was held
to answer more detailed questions about the proposal and to discuss
the rôle of the CKM-LWG.  Points raised during this meeting which
were not covered above are now briefly reviewed.

An important issue that was raised concerned the procedure by which
the groups will reach a consensus on the different quantities studied,
given the sometimes large systematic uncertainties and the prejudices
that members may have regarding different possible approaches. It was
suggested that each group should set up the rules by which decisions
will be made before any final discussion about the summary results
begins. The importance of having the process that leads to any
conclusion public and open to scrutiny, to ensure that
the people involved are accountable, was re-iterated. It was also
emphasized that results should be archived to further promote
accountability.

Another concern, already expressed during the morning session, was
that citations for results would go to the CKM-LWG instead of to the
original papers on which these results are based. To attenuate this
problem, it was suggested that the CKM-LWG could provide on the web
site, next to the various results presented, ready-made \LaTeX\
bibliographic entries of the form: ``CKM-LWG 200x, A. Aardvark {\em et
al.} {\tt (http://www.cpt.univ-mrs.fr/ldg/)}, obtained using the
results from $\ldots$'', with a complete list of the works used.

Finally, the idea of having three coordinators per group was
upheld. However, since the members of the groups were not all present
at the meeting, it was decided that one person from each group would
be responsible for making sure that the coordinators get appointed:
Vittorio Lubicz for the group on quark masses, Anastassios Vladikas
for the group on $B_K$ and Hartmut Wittig for the one on $B$-meson
mixing. It was further suggested that every member in a group should
review all of the literature pertaining to the quantity under study.

%\acknowledgement
\medskip

L.L. thanks J. Christensen for sharing the notes which he took during
the evening discussion at M.I.T. and L. Giusti for discussions about
the project.

\end{document}